# Axiomatic Formulation of the Optimal Transaction Cost Theory in the Legal Process through Cobb-Douglas Optimization

Kwadwo Osei Bonsu,

*k.oseibonsu@pop.zjgsu.edu.cn*

*Doctoral Candidate,*

*School of Economics, School of Law and Intellectual Property, Zhejiang Gongshang University, Hangzhou, Zhejiang, China*

*July, 2021*

**Abstract**

In a legal dispute, parties engage in a series of negotiations so as to arrive at a reasonable settlement. The parties need to present a fair and reasonable bargain in order to induce the WTA (willingness to accept) of the plaintiff and the WTP(willingness to pay) of the defendant. Cooperation can only be attained when the WTA of the plaintiff is less than or equal to the WTP of the defendant, thus WTA≤WTP. From an economic perspective, the legal process can considered as market place of buying and selling claims. High transaction costs decrease the reasonable bargain, thereby making cooperation more appealing to the defendant. On the other hand, low or zero transaction costs means the reasonable bargain is only dependent on the expected gain from winning at trial and the settlement benefit thereby making cooperation more appealing to the plaintiff. Hence, we need to find a way of adjusting the number of settlements with the number of trials in order to maximize the social utility value. This paper uses Cobb-Douglas optimization to formulate an optimal transaction cost algorithm considering the confinement of a generalized legal framework.

**Introduction**

In our previous paper titled '*Mechanism of Instrumental Game Theory in the Legal Process via Stochastic Options Pricing Induction*' (Osei Bonsu and Chen 2020), we deduced that the reasonable bargain can be represented as a function of the expectation benefit and the transaction costs as follows;

$$R_B = \frac{1}{2}[pW_B + S_B] - \frac{1}{2}[C_a + 3C_b]$$

where $R_B$ is the reasonable bargain, p is the probability of plaintiff winning at trial, $W_B$ is the winning benefit from trial for plaintiff, $C_a$ is the administration costs, $C_b$ is the bargaining costs.

We explained that transaction costs comprise of two main parts; $C_b$, bargaining costs(including negotiations, lawyer fees, information exchange etc.) and $C_a$, administration costs(trial costs, including court fees, lawyer fees, time etc.).

Assuming the parties involved are rational, there are four possible scenarios which can occur if the thresholds of the transaction costs are labeled as high or low ;

1) High $C_b$ + High $C_a$

In this case overall transaction costs are too high so the plaintiff will choose trial over settlement since the expectation at trial is greater than bargaining benefit.

2) High $C_b$ + Low $C_a$

In this case the plaintiff will definitely choose trial since it is obvious that the overall benefit from trial is greater than settlement.

3) Low $C_b$ + High $C_a$

In this case the plaintiff will choose settlement if the expectation at trial minus $C_a$ is less than settlement minus $C_b$ .

4) Low $C_b$ + Low $C_a$

In this case the plaintiff will still choose trial because the expectation at trial is greater than settlement.

So we can see that only in the third case will the plaintiff consider settlement. All the other cases will lead to trial thereby increasing the number of trials, which is not the optimal outcome for society.

In the third scenario however, though more cases result in settlements, another problem arises.

To explain the problem, first lets look at the Learned Hands Rule (Kim 2013); this is an algebraic formula, simply written as; B = PL , according to which liability depends on the relation between investment in precaution, B, the product of the probability, P and magnitude of harm, L resulting from the harm caused by the defendant. If PL exceeds B, then the defendant should be liable.

If more cases end up in settlements, more injurers will exercise less precaution since they know there is not much to loose if the case is filed. This will result in more injuries, then more cases. More cases means more trials even though a lower percentage of cases end up in trials, the aggregate number of trials increases, which is neither a socially optimal outcome.

Hence we can infer that the the transaction cost theory in the legal process is not as straightforward as previously discussed in law and economics literature. This is why a more rigorous approach is needed to find a way to estimate not necessarily high or low but an optimal transaction cost algorithm whose parameters and superficial assumptions can be adjusted based on individual cases.

In this paper we will show the effect of administration costs on the outcomes of a legal dispute and propose a methodology for estimating the optimal transaction cost in the legal process in order to obtain a social optimum.

## 1 The set of strategies in a Cooperative legal game

The players in a cooperative legal game jointly choose a set of moral codes m from a set of all possible codes M (Flores Borda 2011),

$$\text{thus } m \in M .$$

The selected moral m contains a set of legal rules R, s.t. R ∈ M, then the players choose a specific legal rule r from R,

$$\text{thus } r \in R .$$

Assuming that the legal rules are supposed to maximize social utility function W according to a particular situation then for the sake of cooperation each player i has to choose strategy $s_i$ from a set of strategies S,

$$\text{thus } s_i \in S.$$

In this case we can find Wmax which is the social maximum. We should note that there is a subset of allowed strategies P(r) under legal rule r within S. So each player i must choose a strategy within P(r),

$$\text{thus } s_i \in P(r) .$$

However we know that in the real world players do not always follow the legal rules. This is why we need to set transaction costs in such a way that the utility U, where $s_i \in P(r)$ will be greater than U, where $s_i \in P(r)'$.

## 2 Utility maximization under constrained strategies

From the $R_B$ equation let $Pc = \frac{1}{2}[pW_B + S_B]$ and $Lc = \frac{1}{2}[C_a + 3C_b]$

where Pc is the expectation benefit component and Lc is the transaction cost component. We can say $R_B + Lc = Pc$. Then the utility function U can be defined as a Cobb-Douglas utility (Shen and Yuming 2020),

$$U(R_B, L_C) = R_B^\alpha L_C^\beta$$

Since we are working within the legal confinement, our feasibility constraint is dependent upon the judgment or the settlement benefit. For a general case, let Lc and $R_B$ be affected by the factor $p_1$ and $p_2$ respectively, we have,

$$p_1 L_C + p_2 R_B \leq P_C$$

The feasibility line has a slope of $-p_1/p_2$, x-intercept of $Pc/p_1$ and y-intercept of $Pc/p_2$.

By Lagrangian approach, we can transform the constrained optimization problem into an unconstrained form. Thus,

$$L = U(R_B, L_C) + \lambda[Pc - p_1 Lc - p_2 R_B]$$

where L is the Lagrange function, $\lambda$ is the Lagrange multiplier.

The resulting first order conditions are;

$$L_{Lc} = \frac{\partial U}{\partial Lc} - \lambda p_1 = 0 \qquad \Rightarrow \qquad \frac{\partial U}{\partial Lc} = \lambda p_1$$

$$L_{R_B} = \frac{\partial U}{\partial R_B} - \lambda p_2 = 0 \qquad \qquad \frac{\partial U}{\partial R_B} = \lambda p_2$$

$$L_\lambda = Pc - p_1 Lc - p_2 R_B = 0 \qquad \qquad Pc = p_1 Lc + p_2 R_B$$

The general solution to the Cobb-Douglas Utility results in;

$$U(L_C, R_B) = L_C^\alpha R_B^\beta$$

Marginal rate of substitution;

$$MRS = \frac{\alpha R_B}{\beta L_C}$$

then the first rule of utility yields,

$$MRS = \frac{p_1}{p_2}$$

$$\Rightarrow \frac{\alpha R_B}{\beta L_C} = \frac{p_1}{p_2} \Rightarrow R_B = \frac{p_1}{p_2} \cdot \frac{\beta}{\alpha} L_C$$

Substituting $R_B$ into the feasibility constraint yields,

$$P_C = p_1 L_C + p_2 (\frac{p_1}{p_2} \frac{\beta}{\alpha} L_C)$$

$$P_C = p_1 L_C + \frac{\beta}{\alpha} p_1 L_C = (\frac{\alpha + \beta}{\alpha}) p_1 L_C$$

$$L_C = (\frac{\alpha}{\alpha + \beta}) \frac{Pc}{p_1} \Rightarrow p_1 = (\frac{\alpha}{\alpha + \beta}) \frac{Pc}{L_C}$$

Likewise we can say that,

$$R_B = (\frac{\beta}{\alpha + \beta}) \frac{Pc}{p_2} \Rightarrow p_2 = (\frac{\beta}{\alpha + \beta}) \frac{Pc}{R_B}$$

Substituting $p_1$ and $p_2$ into the resulting first order conditions we get,

$$\frac{\partial U}{\partial Lc} = \lambda (\frac{\alpha}{\alpha + \beta}) \frac{Pc}{Lc}$$

$$\frac{\partial U}{\partial R_B} = \lambda (\frac{\beta}{\alpha + \beta}) \frac{Pc}{R_B}$$

$\because U(L_C, R_B) = L_C^\alpha R_B^\beta$ first order partial derivatives result in;

$$\frac{\partial U}{\partial Lc} = \lambda(\frac{\alpha}{\alpha+\beta})\frac{Pc}{Lc} = \alpha L_c^{\alpha-1} R_B^{\beta}$$

$$\frac{\partial U}{\partial R_B} = \lambda(\frac{\beta}{\alpha+\beta})\frac{Pc}{R_B} = \beta L_c^{\alpha} R_B^{\beta-1}$$

$$\Rightarrow (\frac{\lambda}{\alpha+\beta})Pc = L_c^{\alpha} R_B^{\beta}$$

$$\therefore U(Lc, R_B) = (\frac{\lambda}{\alpha+\beta})Pc$$

For Karush-Kuhn-Tucker conditions to hold, lambda must be greater than zero.
In the legal negotiation zero lambda means zero utility which is not acceptable for a rational player (Felipe; Jesus and Adams 2005). So lambda must always be greater than zero.
Meaning when $Lc$ gets bigger we need to adjust $R_B$ to maintain at least a positive utility. The easiest way to do that is to increase $Pc$, the expectation benefit. Thereby increasing either $W_B$ or $S_B$.

When Lc, the transaction cost gets lower, lambda gets bigger thereby increasing the utility of the players, which is in conformity with Coase's Theorem.
However, when $Lc$ is very low, say close to zero, lambda is nearly totally dependent on $R_B$ which also gets bigger due to the absence of or very low transaction costs. So lambda gets smaller resulting in lower utility.
Therefore we can say that transaction costs cannot be too low nor too high.

### 3 Optimal transaction cost algorithm

In a legal dispute if we hold the attorney fees and other costs $C_b$ in the reasonable bargain equation constant, then we can say that a function Φ of the transaction costs is dependent on $Lc$.

Let $Li = (L_1, \ldots, L_n)^T$, then $R_B + \Phi(Lci) \leq Pc$

$$\Rightarrow \phi(Lci) = \sum_{i=1}^{n} \phi i(Li)$$

As shown in the previous section,

$$\phi(Lci) \neq 0; \phi(Lci) > 0; \phi(Lci) < R_B$$
$$\Rightarrow 0 < \phi(Lci) < R_B$$

A more realistic model of the transaction cost in the justice system is that it should be proportional to the intensity of the case.

Let $\alpha_i^+$ and $\alpha_i^-$ be the cost rates associated with $Li=(L_1,.......L_n)^T$, we can say that,

$$\phi i(Li) = \begin{cases} \alpha_i^+ Li, Li > 0 \\ -\alpha_i^- Li, Li < 0 \\ 0, Li = 0 \end{cases}$$

If we consider the fixed costs $C_b$ we have;

$$\phi i(Li) = \begin{cases} C_b + \alpha_i^+ Li, Li > 0 \\ C_b - \alpha_i^- Li, Li < 0 \\ 0, Li = 0 \end{cases}$$

$$\Rightarrow \phi i(Li) = \begin{cases} 0, Li = 0 \\ C_b + \alpha_i \mid Li \mid, Li \neq 0 \end{cases}$$

Lambda from the Lagrangian equation describes the shadow values of the feasibility constant Pc.

$$\Rightarrow U(Lci, R_B) = (\frac{\lambda i}{\alpha + \beta})\left[\left(\sum_{i=1}^{n} \phi i(Li)\right) + R_B\right]$$

The determinant of the bordered Hessian can be deduced as;

$$\det(H) = \begin{vmatrix} 0 & -\lambda p_1^* & -\lambda p_2^* \\ -\lambda p_1^* & \frac{\partial^2 L}{\partial L_C^2} & \frac{\partial^2 L}{\partial L_C \partial R_B} \\ -\lambda p_2^* & \frac{\partial^2 L}{\partial L_C \partial R_B} & \frac{\partial^2 L}{\partial R_B^2} \end{vmatrix}$$

In terms of lambda and the feasibility constraint we have,

$$\det(H) = \begin{vmatrix} 0 & -\lambda p_1^* & -\lambda p_2^* \\ -\lambda p_1^* & -\lambda(\frac{\alpha}{\alpha+\beta})\frac{Pc}{L_C^2} & 0 \\ -\lambda p_2^* & 0 & -\lambda(\frac{\beta}{\alpha+\beta})\frac{Pc}{R_B^2} \end{vmatrix}$$

resulting in,

$$\|H\| = 0 - (-\lambda p_1^*)\left[(-\lambda p_1^*)\left(-\lambda(\frac{\beta}{\alpha+\beta})\frac{Pc}{R_B^2}\right)\right] + (-\lambda p_2^*)\left[-(-\lambda p_2^*)\left(-\lambda(\frac{\alpha}{\alpha+\beta})\frac{Pc}{L_C^2}\right)\right]$$

In terms of the change in Lc and R$_B$ from the utility function we have,

$$\det(H) = \begin{vmatrix} 0 & -\lambda p_1^* & -\lambda p_2^* \\ -\lambda p_1^* & \alpha(\alpha-1)L_C^{\alpha-2}R_B & 0 \\ -\lambda p_2^* & 0 & \beta(\beta-1)L_C^\alpha R_B^{\beta-2} \end{vmatrix}$$

resulting in;

$$\|H\| = 0 - (-\lambda p_1^*)\left[(-\lambda p_1^*)(\beta(\beta-1)L_C^\alpha R_B^{\beta-2})\right] + (-\lambda p_2^*)\left[-(-\lambda p_2^*)(\alpha(\alpha-1)L_C^{\alpha-2}R_B)\right]$$

From the two equations resulting from det(H), we can see consistent sign ++ or -- so in order to obtain a local maxima we need a set of values of alpha that will make sure that the signs are consistent.

If $\alpha_p=(\alpha_1,\ldots,\alpha_n)$ and $\beta_q=(\beta_1,\ldots,\beta_n)$ s.t. $\alpha \epsilon \alpha_p$ and $\beta \epsilon \beta_q$, then;

$$\exists \alpha^* s.t. \left(\frac{\lambda i}{\alpha+\beta}\right) > 0$$

in order to make det(H)>0.

Therefore this particular alpha which is a power of Lc in the utility function U(Lc, $R_B$) leads to the measure of the optimal transaction cost.

Thus $$U^*(L^*c, R_B) = \left(\frac{\lambda^*}{\alpha^*+\beta}\right)\left[L^*c + R_B\right]$$

**Conclusion**
In this paper, we first discussed two big problems that arise when we label specific thresholds of transaction costs in the legal process as previously done by law and economics scholars. The first problem is that high transaction costs may induce cooperation while low transaction costs may induce non-cooperation which is in contrast with Coase's Theorem which says that lower transaction costs induce private agreements. The second problem is that since the reasonable bargain is built around the threat value, it does create a sense of fairness(i.e restitution plus projected payoff) to the parties involved. However, since our initial claim is that the legal process is a market, we need to put a price on the claim rather than just a reasonable bargain. This means the reasonable bargain is just the minimum mutually acceptable cooperative position.
We have shown how transaction costs influence the set of strategies played by players in a legal game, described the essence of a social utility function and how it could be maximized through an optimal transaction cost algorithm. We have also used the Cobb-Douglas optimization to show how an optimal transaction cost algorithm can be deduced using the Lagrangian approach.
In most court cases, fees are intended to shift the costs of the justice system from taxpayers to litigants who are seen as the 'users' of the courts. These costs may be used for covering parts of the justice process such as court-appointed attorney fees, court clerk fees, forensic fees, equipment fees, utility etc. In this research we recommend that the courts should balance their costs between administration costs and taxpayers support for optimal distribution of social welfare.

Further research is needed to determine a more specific transaction cost algorithm based on empirical evidence or an econometric approach to complement the theoretical foundations derived in this paper. Finally, there is the need for further research into how psychological factors affect the probability of winning at trial and the link between transaction costs and behavioral patterns in the legal process.

**Conflicts of Interests:**

There are no conflicts of interests in this paper

**Tables and Figures:**

There are no tables or figures in this paper